\documentclass{article}
\usepackage{todonotes}
\usepackage{graphicx}
\usepackage{algorithm}
\usepackage{algpseudocode}

\newcommand{\REDACT}[1]{$\Box REDACTED \Box$} 

\newcommand{\redactCollege}[1]{[a U.S. University]}  

\newcounter{boldifyCounter}
\newcounter{fixmeSectionCounter}
\newcounter{fixmeTotalCounter}
\makeatletter
\@addtoreset{fixmeSectionCounter}{section}
\@addtoreset{fixmeSectionCounter}{subsection}
\@addtoreset{boldifyCounter}{section}
\@addtoreset{boldifyCounter}{subsection}
\makeatother

\newcommand{\boldify}[1]{}
\ifdefined\boldifyON
	\renewcommand{\boldify}[1]{
        \par\noindent
		\stepcounter{boldifyCounter}
		\textbf{{\color{green}**}
		~\arabic{section}.\arabic{subsection}.\arabic{boldifyCounter}
		: #1} 
	}
\fi 

\newcommand{\reportOnFIXME}{%
    \newcount\iterCounter
    \iterCounter=1
    \newcount\endCounter
    \endCounter=\totvalue{fixmeTotalCounter}
    \advance \endCounter +1
    There are 
    {\color{red}\total{fixmeTotalCounter}} 
    FIX\_ME\\
    links:
    \loop
        \hyperlink{fixTag\the\iterCounter}{\#\the\iterCounter}
        \advance \iterCounter +1
    \ifnum \iterCounter < \endCounter
    \repeat
}

\newcommand{\FIXME}[1]{} 
\ifdefined\fixmeON
	\renewcommand{\FIXME}[1]{\par\noindent
		\stepcounter{fixmeSectionCounter}\stepcounter{fixmeTotalCounter}
		{\color{red}\fbox{\color{black}
			\parbox{.965\linewidth}{
				\textbf{\hypertarget{fixTag\thefixmeTotalCounter}{FIXME}	\arabic{section}.\arabic{subsection}.
        		\arabic{fixmeSectionCounter} (\color{red}
        		\#\arabic{fixmeTotalCounter}):} #1}}
        }
	}
\fi 

\newcommand{\FIXED}[1]{}
\ifdefined\fixmeON
	\renewcommand{\FIXED}[1]{\par\noindent%
		{\color{black}\fbox{\color{black}%
			\parbox{.99\columnwidth}{%
				\color{blue}#1}}%
        }
	}
\fi 

\newcommand{\draftStatus}[1]{}
\ifdefined\draftStatusON
	\renewcommand{\draftStatus}[1]{
        \hfill **#1
	}
\fi 

    \PassOptionsToPackage{numbers, compress}{natbib}

   \usepackage[final]{neurips_2024}

\usepackage[utf8]{inputenc} 
\usepackage[T1]{fontenc}    
\usepackage{hyperref}       
\usepackage{url}            
\usepackage{booktabs}       
\usepackage{amsfonts}       
\usepackage{nicefrac}       
\usepackage{microtype}      
\usepackage{xcolor}         

\usepackage{amsmath}
\usepackage{graphicx}
\usepackage{array}
\usepackage{longtable}
\usepackage{tabularx} 

\title{
A Formal Framework for Assessing and Mitigating Emergent Security Risks in Generative AI Models: Bridging Theory and Dynamic Risk Mitigation
}

\author{%
    Aviral Srivastava \\
    College of IST\\
    Pennsylvania State University\\
    State College, PA 16802\\
    \texttt{aks7873@psu.edu} \\
    \AND
    Sourav Panda \\
    College of IST\\
    Pennsylvania State University\\
    State College, PA 16802\\
    \texttt{sbp5911@psu.edu} \\
}

\begin{document}

\maketitle

\begin{abstract}
As generative AI systems, including large language models (LLMs) and diffusion models, advance rapidly, their growing adoption has led to new and complex security risks often overlooked in traditional AI risk assessment frameworks. 
This paper introduces a novel formal framework for categorizing and mitigating these emergent security risks by integrating adaptive, real-time monitoring, and dynamic risk mitigation strategies tailored to generative models' unique vulnerabilities. We identify previously under-explored risks, including latent space exploitation, multi-modal cross-attack vectors, and feedback-loop-induced model degradation. Our framework employs a layered approach, incorporating anomaly detection, continuous red-teaming, and real-time adversarial simulation to mitigate these risks. We focus on formal verification methods to ensure model robustness and scalability in the face of evolving threats. 
Though theoretical, this work sets the stage for future empirical validation by establishing a detailed methodology and metrics for evaluating the performance of risk mitigation strategies in generative AI systems. This framework addresses existing gaps in AI safety, offering a comprehensive road map for future research and implementation.
\end{abstract}

\section{Introduction}
Generative AI models, including large language models (LLMs) and multimodal systems like diffusion models, have dramatically transformed a variety of industries. These models enable advanced capabilities in content creation, natural language understanding, and image generation, yet their complex architectures present substantial and often unforeseen security vulnerabilities. The rapid deployment of generative AI systems has far outpaced the development of comprehensive security frameworks, leaving critical gaps in our ability to identify and mitigate risks. Traditional AI safety approaches, while effective in identifying issues like adversarial attacks and data leakage, are insufficient to manage the dynamic, emergent threats specific to generative AI.

Current risk assessment models fall short in addressing the evolving nature of these systems, particularly their susceptibility to latent space exploitation, adversarial prompt manipulation, and feedback-loop-induced degradation. Moreover, the integration of multimodal capabilities introduces novel attack surfaces, allowing adversaries to exploit vulnerabilities across text, image, and video simultaneously. These emergent risks pose significant threats to both the functionality of generative AI systems and the broader ecosystem in which they operate, particularly in fields like healthcare, finance, and security.
This paper seeks to address these gaps by proposing a formal framework for assessing and mitigating emergent security risks in generative AI models. To guide the development and evaluation of this framework, we pose the following research questions (RQs):
\begin{itemize}
    \item \textbf{RQ1:} How can emergent security risks in generative AI models, such as latent space exploitation and multimodal cross-attack vectors, be systematically identified and categorized across the model lifecycle?
    \item \textbf{RQ2:} What adaptive, real-time monitoring techniques can be integrated into generative models to provide continuous risk assessment and dynamic mitigation in response to evolving threats?
    \item \textbf{RQ3:} How can formal verification methods be utilized to ensure the robustness and scalability of generative AI models against emergent threats while maintaining system efficiency?
\end{itemize}

This paper presents a layered risk identification approach that spans the entire lifecycle of generative models, from data ingestion to model inference, incorporating continuous adversarial testing, anomaly detection, and dynamic watermarking techniques. While this framework is theoretical, it offers a robust foundation for future empirical validation and sets forth a roadmap for the development of scalable, adaptive security measures in generative AI.

By addressing both the theoretical underpinnings and practical challenges of securing generative AI models, this paper aims to fill critical gaps in AI safety research. Our framework not only enhances existing methods but also introduces novel solutions designed to evolve alongside the generative models they protect, ensuring their secure deployment in increasingly complex environments.

\section{Literature Review}

\paragraph{Security Risks in Generative AI}
Generative AI systems, such as large language models (LLMs) and diffusion models, present unique security challenges due to their complex architectures and emergent capabilities. Barrett et al. discuss the dual-use dilemma of generative AI, highlighting both its capabilities and the security risks it poses. 
They synthesize findings from a workshop held at Google, co-organized by Stanford University and the University of Wisconsin-Madison, emphasizing the need for robust security measures~\cite{barrett2023identifying}. Similarly, Humphreys et al. examine the ethical obligations companies have when implementing generative AI, focusing on the potential cyber security risks and the moral responsibilities involved~\cite{humphreys2024ai}. Feffer et al. explore the concept of red-teaming in generative AI, analyzing its effectiveness and limitations in identifying and mitigating security risks~\cite{feffer2024red}.

\paragraph{Risk Assessment in AI}
Risk assessment is a critical component of AI governance. Novelli et al. propose a methodology for assessing AI risk magnitudes, focusing on the construction of real-world risk scenarios and applying a proportionality test to balance competing values~\cite{novelli2024ai}. Schmitz et al. provide a systematic overview of risk aggregation schemes used in existing AI risk assessment frameworks, focusing on how trade-offs among risk dimensions are incorporated~\cite{schmitz2024global}.

\paragraph{Risk Mitigation in AI}
Effective risk mitigation strategies are essential for securing AI systems. 
Kaminski describes the growing convergence around risk regulation in AI governance, offering an analytic framework for understanding the use of risk regulation in AI governance~\cite{Kaminski2023Regulating}. Parimala provides a comprehensive overview of recent advances in anomaly detection techniques, highlighting the role of AI and machine learning in this field~\cite{parimala2024introductory}.

\paragraph{AI Safety}
AI safety is a crucial aspect of AI development and deployment. Shah and Mishra review the advancements in AI-driven occupational health and safety technologies, offering predictive insights and risk mitigation strategies~\cite{shah2024artificial}. Deepak discusses the global enthusiasm around AI safety, highlighting the challenges and potential issues associated with current AI safety measures~\cite{deepak2024ai}.

\paragraph{Multi-Modal Evaluation in AI}
Multi-modal evaluation is essential for assessing the performance of AI systems across different modalities. Chen et al. introduce a comprehensive benchmark for evaluating large vision-language models in medical applications, addressing the limitations of current benchmarks~\cite{chen2024gmai}.

\paragraph{Dynamic Watermarking in AI}
Dynamic watermarking is a promising technique for protecting the integrity of AI-generated content. Zhong et al. present a comprehensive overview of deep learning-based image watermarking techniques, focusing on robustness and adaptability~\cite{zhong2023brief}.

\paragraph{Real-Time Anomaly Detection in AI}
Real-time anomaly detection is critical for identifying and mitigating security threats in AI systems. Dini and Saponara explore the design and assessment of real-time anomaly detection techniques for automotive cybersecurity, leveraging neural networks and fingerprinting techniques~\cite{dini2023design}.
\paragraph{}
While the existing body of work provides valuable insights into the security risks, risk assessment methods, and mitigation strategies for AI systems, they fall short in addressing the dynamic and evolving nature of generative AI models. Current methods often lack the ability to continuously adapt to new, unforeseen vulnerabilities that arise post-deployment. Our proposed framework addresses these gaps by integrating continuous risk monitoring, real-time adversarial simulations, and adaptive mitigation techniques that evolve alongside the AI models. This dynamic approach ensures that emergent threats are detected and mitigated in real-time, providing a more robust and scalable solution compared to static, pre-deployment assessments.

\section{Novel Security Risks in GenAI}

\subsection{Introduction of New Vulnerabilities:}
While current research has identified some key vulnerabilities in generative AI (GenAI) systems, such as hallucinations, data leakage, and adversarial prompt manipulation, emerging threats have not been fully explored. In this section, we analyze a set of novel security risks, focusing on unsupervised learning dynamics, self-improving models, latent space exploitation, and cross-modal attacks in multimodal systems. These vulnerabilities represent critical areas for future research, particularly in terms of adaptive, real-time risk monitoring.

\subsection{Unsupervised Learning Vulnerabilities}
\paragraph{Risk Overview:}
Generative AI models, particularly those trained on unsupervised or self-supervised datasets, rely heavily on vast, uncurated data sources scraped from public repositories. Without the explicit labeling present in supervised learning, these systems are prone to absorbing noisy, biased, or harmful data. As models continue to evolve based on this data, they may reinforce incorrect associations or harmful biases, often without detection. This risk is exacerbated when feedback loops emerge between model outputs and training data.
\paragraph{Potential Impact:}
\begin{itemize}
    \item \textbf{Bias Amplification:} Generative models trained on unsupervised datasets may inadvertently learn and perpetuate societal biases, including racial, gender, or cultural biases. The absence of real-time oversight makes these biases ingrained, resulting in systemic output biases.
    \item \textbf{Feedback Loops:} When generative models interact with user-generated content on platforms like social media, a feedback loop can develop. In this cycle, the biased or harmful outputs generated by the model are reintroduced as new data, further reinforcing incorrect or problematic patterns. Over time, this can degrade the model's accuracy and increase its vulnerability to adversarial manipulation.
\end{itemize}

\textbf{Example Case:}
A GenAI model trained on uncurated social media data may generate biased outputs when asked to summarize trending topics. If the model then scrapes its own generated content as part of future training data, it perpetuates and amplifies the biases present in the original input~\cite{barrett2023identifying, chua2024ai}

\subsection{Self-Improvement Risks}
\paragraph{Risk Overview:}
One of the most exciting but risky developments in GenAI is the emergence of self-improving models. These systems are designed to fine-tune themselves in real time based on feedback from their interactions. While this self-adaptive capability offers efficiency gains, it also introduces the risk of unchecked bias reinforcement and the propagation of security vulnerabilities. Without human oversight, these models may learn from harmful inputs, deepening their vulnerabilities over time.
\paragraph{Potential Impact:}
\begin{itemize}
    \item \textbf{Bias Reinforcement:} Self-improving models that adjust based on real-world data without external oversight may unintentionally reinforce harmful biases. 
    Over time, small vulnerabilities could become more severe as the model's weights and parameters shift to accommodate flawed or adversarial inputs
    \item \textbf{Exacerbating Security Issues:} Without structured validation, self-improving models could become increasingly susceptible to adversarial prompts. 
    Malicious users might feed the model adversarial or biased inputs, which the model could learn from, thereby worsening its vulnerabilities in subsequent interactions.
\end{itemize}

\textbf{Example Case:}
A self-improving chatbot that interacts with malicious users may start learning from harmful, biased, or adversarial inputs, causing it to produce increasingly flawed or harmful outputs. As the chatbot fine-tunes itself based on these interactions, it becomes more vulnerable to adversarial attacks, with its security issues compounding over time~\cite{chua2024ai, SistoHalm2024}.

\subsection{Hidden Information Leakage: Latent Space Vulnerabilities}
\paragraph{Risk Overview:}
One of the less-discussed but critical risks in generative models is latent space exploitation. 
The latent space represents the internal high-dimensional feature representations learned by the model. While latent spaces are crucial for model functionality, they can inadvertently retain sensitive patterns from the training data. Attackers with knowledge of the model's architecture can exploit these latent spaces to extract sensitive information, leading to significant privacy concerns in sectors like healthcare and finance.
\paragraph{Potential Impact:}
\begin{itemize}
    \item \textbf{Latent Space Exploitation:} Attackers with sophisticated knowledge of a model’s architecture could manipulate inputs to probe the latent space and retrieve sensitive information embedded in these internal representations. 
    For example, in healthcare models, the latent space might reveal sensitive patient data even when the output is anonymized.
    \item \textbf{Inferred Privacy Leaks:} Through adversarial probing, an attacker could generate outputs that reveal latent correlations, potentially identifying individuals from supposedly anonymized data or uncovering proprietary business strategies from models trained on confidential data.
\end{itemize}

\textbf{Example Case:}
In financial systems, attackers could probe the latent space of a generative model used for stock prediction to uncover sensitive market trends or proprietary investment strategies, leading to significant financial or privacy breaches~\cite{barrett2023identifying, chua2024ai}

\subsection{Simultaneous Multimodal Exploitation}
\paragraph{Risk Overview:}
Multimodal generative models that process and generate text, images, audio, and video simultaneously present new attack surfaces. These models integrate multiple types of data, making them vulnerable to cross-modal exploits. An attack targeting one modality (e.g., text) could expose weaknesses in another modality (e.g., images or videos), enabling adversaries to exploit the model across different dimensions simultaneously.
\paragraph{Potential Impact:}
\begin{itemize}
    \item \textbf{Cross-Modal Consistency Exploits:} Attackers can inject adversarial inputs into one modality (e.g., an image) that lead to harmful outputs in another modality (e.g., misleading or biased text). 
    For instance, an adversarial image might be paired with a deceptive or biased textual description, which reinforces false or harmful narratives.
    \item \textbf{Coordinated Multimodal Attacks:} Attackers could simultaneously manipulate multiple modalities to generate realistic yet harmful content, such as deepfakes. 
    This is particularly concerning in scenarios like news generation or multimedia content creation, where both visual and audio elements can be manipulated to spread misinformation or discredit individuals.
\end{itemize}

\textbf{Example Case:}
In a news generation scenario, an adversary could generate fake articles supported by AI-generated deepfake images or videos. These multimodal attacks make it more challenging for users to distinguish between legitimate and fake content, amplifying the potential for misinformation~\cite{barrett2023identifying, chua2024ai, SistoHalm2024}.

\section{Parameters for Risk Categorization}
Based on the insights from NIST’s Risk Management Framework (RMF) and the ISO 27001 cybersecurity standards~\cite{PrivacyEngine2023ISOvsNIST, NIST2024CyberRiskScoring}, we can classify risks using the following parameters

\paragraph{Potential Impact}
\begin{itemize}
    \item Low: Minimal consequences if exploited, affecting a limited number of users or non-critical processes. 
    For instance, generating biased but non-malicious text in an isolated scenario.
    \item Medium: Significant impact on some operations or users, such as privacy breaches affecting identifiable groups.
    \item High: Severe consequences, such as critical system failures or widespread misinformation, impacting infrastructure, security, or public trust.
\end{itemize}

\paragraph{Exploitability}
\begin{itemize}
    \item Low: Requires a sophisticated attack vector or rare conditions, making exploitation unlikely.
    \item Medium: Moderate effort or specific conditions required for an exploit, such as known vulnerabilities in handling adversarial inputs
    \item High: Easily exploitable due to well-known vulnerabilities or lack of adequate safeguards, such as unsupervised learning models prone to bias without regular updates.
\end{itemize}
\paragraph{Scope of Effect}
\begin{itemize}
    \item Low: Localized effects with minimal spread, possibly confined to a single user or output.
    \item Medium: Broader reach, affecting multiple systems, users, or outputs.
    \item High: Extensive impact, potentially leading to cascading failures across multiple systems, modalities, or platforms.
\end{itemize}

By combining these parameters, we can systematically assess the severity of each security risk in generative AI models. Refer to the Appendix for a table to categorize the key risks outlined in Section 3 of the paper.

\section{Novel Risk Mitigation Strategies}
We propose a series of novel risk mitigation strategies for generative AI models, focusing on adaptive mechanisms that evolve in response to new attack patterns. These strategies include real-time filtering, dynamic watermarking, multi-layered mitigation, and human-AI collaboration.
\paragraph{Contextual Real-Time Filtering}
We introduce a real-time filtering system that evolves based on new attack patterns. For example, a filtering mechanism could detect and block problematic prompts (e.g., adversarial input) before they are processed, learning from each failed attack.

\paragraph{Dynamic Watermarking}
We propose the development of a self-evolving watermarking system, where watermark techniques adapt as adversaries develop new removal methods. The watermarking algorithm can evolve based on detection failures \cite{barrett2023identifying}.

\paragraph{Multi-Layered Mitigation}
We propose a multi-tier system for mitigating risks at various stages of the model lifecycle:
\begin{itemize}
    \item \textbf{Tier 1:} Pre-deployment filtering, applied during model training to ensure that adversarial inputs are detected and mitigated before deployment.
    \item \textbf{Tier 2:} Inference-time defense, focusing on real-time output filtering to detect and block potentially harmful outputs.
    \item \textbf{Tier 3:} Post-inference validation, where high-risk interactions are validated after generation to reduce false positives and mitigate any remaining risks.
\end{itemize}

\paragraph{Human-AI Collaboration}
In decision-critical environments, we suggest mechanisms for human oversight. Feedback loops can be incorporated where human analysts intervene when GenAI outputs are flagged as potentially harmful. This collaboration between human oversight and AI-driven analysis will ensure that higher-risk outputs receive necessary validation and correction when required.

Refer to Appendix Figure 1 for a visualization of the Working Structure of the Formal Framework for Risk Assessment in Generative AI

\section{Formal Framework for Risk Assessment}
In this section, we present a comprehensive formal framework for systematically assessing and mitigating security risks in generative AI models. This framework leverages formal verification, continuous monitoring, and adaptive risk mitigation to address both known and emergent threats. By adopting a layered approach, we ensure that risks are identified and mitigated at critical stages of the GenAI lifecycle, from data ingestion to model inference.

\subsection{Theoretical Grounding}
The framework is grounded in formal verification methods, which have been widely used in software engineering and AI safety to mathematically guarantee that a system behaves according to its specifications \cite{PrivacyEngine2023ISOvsNIST, NIST2024CyberRiskScoring}. Formal verification is particularly suitable for assessing generative AI models because of their complexity and the emergent behaviors they exhibit. These methods allow us to model expected behavior and prove that deviations—such as biases, adversarial attacks, or data leakage—will be detected and mitigated before deployment.

\subsection{Risk Identification Layers}
Our framework operates through a layered approach, designed to identify and mitigate risks at three key stages of the generative AI lifecycle: the data layer, model training layer, and model inference layer. Each stage involves distinct methodologies for risk identification, leveraging formal verification, adversarial testing, and real-time anomaly detection.

\subsubsection{Data Layer}
At the data layer, risks primarily involve biases, noisy data, and privacy violations that could influence model behavior later in the lifecycle. This layer applies formal verification techniques to ensure that the data used for training is both consistent and free from harmful biases or inaccuracies.

\textbf{Formal Verification Algorithm:} We deploy a bias filtering algorithm that uses mathematical models to verify that the training data does not reinforce societal biases, such as gender or racial biases. This approach detects anomalies within the dataset, ensuring that harmful patterns are flagged before they can affect model outputs. This algorithm functions by analyzing patterns in the data that deviate from established baselines of fairness and accuracy \cite{NIST2024CyberRiskScoring}.

\textbf{Dynamic Data Monitoring:} In addition to formal verification, continuous monitoring is deployed to detect new patterns of bias that may emerge in real time. A real-time anomaly detection system monitors the data for unexpected shifts in distribution or content. If a bias or noisy data is detected, the data can be flagged, and the model training process is halted for further review.

\subsubsection{Model Training Layer}
During model training, adversarial manipulation becomes a significant threat, where attackers may attempt to perturb the training data to manipulate the model’s learning. Our framework integrates automated adversarial testing directly into the training process, enabling the system to continuously probe for weaknesses before the model is deployed.

\textbf{Adversarial Testing Module:} This module employs real-time adversarial simulation that introduces adversarial inputs at various points during the training process to test the model's robustness. The simulation leverages adversarial perturbations—slight modifications to the input data that are designed to mislead the model. These perturbations mimic real-world attacks, allowing the framework to detect vulnerabilities before deployment. The module monitors how effectively the model resists adversarial inputs and adjusts accordingly.

\textbf{Metrics for Robustness:} Metrics used to evaluate the effectiveness of adversarial testing include the model’s accuracy under adversarial conditions, the degree of perturbation required to mislead the model, and the model’s ability to maintain consistency in outputs across multiple adversarial inputs. These metrics help define what constitutes a "high-risk" versus "low-risk" vulnerability. For example:
\begin{itemize}
    \item \textbf{High-risk:} If a small perturbation results in significantly misleading outputs, indicating that the model is highly susceptible to adversarial attacks.
    \item \textbf{Low-risk:} If the model remains robust even under large adversarial perturbations, demonstrating strong resistance to manipulation.
\end{itemize}

\subsubsection{Model Inference Layer}
At the inference stage, risks are primarily associated with real-time attacks and unforeseen vulnerabilities that manifest when the model interacts with external environments. This is where dynamic risk monitoring becomes critical.

\textbf{Real-time Adversarial Simulation and Response:} During inference, each input-output pair is subjected to real-time adversarial simulation. This system continuously analyzes the model’s interactions with external inputs, looking for any signs of adversarial activity. For instance, when the model encounters a prompt, the adversarial simulation generates variations of the input designed to probe for weaknesses (e.g., modifying a prompt to generate biased or harmful outputs). These adversarial tests run concurrently with the model’s normal operation, ensuring that vulnerabilities can be detected and responded to in real-time \cite{PrivacyEngine2023ISOvsNIST}.

\textbf{Latent Space Monitoring:} Anomaly detection in the latent space is employed to track patterns that emerge deep within the model’s architecture. By establishing a statistical baseline of what constitutes "normal" latent space activity, the framework can detect deviations that suggest potential risks, such as information leakage or bias amplification. The system continuously monitors latent space activity for anomalies, which could indicate an ongoing attack or emergent bias that was not detected during training \cite{NIST2024CyberRiskScoring}.

\textbf{Dynamic Watermarking:} To further secure the inference layer, the framework implements dynamic watermarking, which involves embedding invisible markers in the outputs generated by the model. These watermarks evolve to resist adversarial removal techniques. If the watermark is compromised or missing, it indicates potential tampering or misuse of the model outputs.

\subsection{Dynamic Risk Monitoring}
To address the continuously evolving threat landscape, our framework incorporates dynamic risk monitoring mechanisms that operate at all stages of the model lifecycle. 
These include:
\begin{itemize}
    \item \textbf{Real-time Adversarial Simulations:} These simulations are continuously applied during both training and inference to detect potential vulnerabilities early and predict the likelihood of exploitation. 
    The feedback loop generated by adversarial testing informs the model’s security posture, dynamically adjusting its architecture as new threats are identified.
    \item \textbf{Anomaly Detection Algorithms:} Anomalies in the model’s performance or latent space activity are flagged in real-time, triggering preemptive countermeasures. 
    These algorithms are particularly focused on detecting latent space leaks—high-dimensional spaces where information from the training data can leak into the model’s predictions.
    \item \textbf{Continuous Model Refinement:} When vulnerabilities are detected, the model is fine-tuned in real time. 
    This process leverages reinforcement learning techniques, where the model autonomously learns from adversarial inputs and updates its parameters to resist similar attacks in the future. 
    This continuous adjustment ensures that the model remains secure even as new threats emerge \cite{NIST2024CyberRiskScoring}.
\end{itemize}

\subsection{Metrics for Evaluating the Framework}
The success of the proposed framework is evaluated through a series of well-defined metrics:
\begin{itemize}
    \item \textbf{Attack Detection Rate:} The percentage of adversarial attacks successfully detected during training and inference.
    \item \textbf{Response Time to Threats:} The time it takes for the system to detect and respond to an identified vulnerability.
    \item \textbf{Model Robustness:} The model’s accuracy and consistency when subjected to adversarial inputs or anomalies.
    \item \textbf{False Positive Rate:} The frequency of incorrectly flagged inputs or behaviors that do not represent actual security risks.
    \item \textbf{Latency Impact:} The effect of continuous monitoring and adversarial simulation on the model’s real-time performance, ensuring that security measures do not compromise system efficiency.
\end{itemize}
These metrics are essential for distinguishing between "high-risk" and "low-risk" scenarios, enabling the model to prioritize mitigation efforts where they are most needed.

Refer to Appendix Figure 2 for a visualization of the Formal Framework for Risk Assessment in Generative AI.

By integrating formal verification, continuous adversarial simulation, and real-time anomaly detection, this framework provides a comprehensive, adaptive approach to securing generative AI models. The framework’s layered architecture ensures that vulnerabilities are identified and mitigated throughout the entire model lifecycle, from deployment training. Though theoretical, this framework lays the foundation for empirical validation and future development in securing generative AI systems against evolving threats.

\section{Limitations}
This paper presents a theoretical framework for assessing and mitigating emergent security risks in generative AI models. While the framework is rooted in existing risk management methodologies such as those from NIST, ISO, and MITRE, there are several inherent limitations:

\paragraph{Absence of Empirical Validation}
As this work is purely theoretical, none of the proposed mitigation strategies have been tested in real-world environments. The effectiveness of dynamic risk monitoring, adversarial simulations, and anomaly detection systems must be empirically validated before they can be widely adopted. The lack of empirical testing means the practical feasibility of the framework remains unproven, especially in diverse AI use cases such as healthcare, finance, and large-scale media generation.

\paragraph{Dynamic Threat Landscape}
The constantly evolving nature of generative AI means that new, unforeseen vulnerabilities could emerge that are beyond the scope of this framework. Models like GPT and other multimodal systems are rapidly changing, and novel forms of attacks or biases may arise that our current framework may not fully address. As these systems scale, managing real-time risk also becomes more complex, and our framework may need to be adapted over time to incorporate future security trends.

\paragraph{Resource and Scalability Constraints}
The continuous risk monitoring and real-time adversarial simulation components of this framework could be resource-intensive. Ensuring that these defenses operate effectively without slowing down model performance or requiring significant computational power poses a real challenge. Small- to medium-sized organizations may find it difficult to implement such resource-heavy solutions.

\paragraph{Latent Space Monitoring}
Monitoring the latent space in real time presents significant technical challenges. While we propose anomaly detection for latent space monitoring, the ability to interpret the high-dimensional features in latent spaces and identify security risks remains an under-explored area. Future research will need to refine this approach and improve the interpretability of latent space representations.

\section{Future Directions}
Despite these limitations, the theoretical framework presented in this paper provides a solid foundation for further research and development in generative AI security. Key avenues for future work include:

\paragraph{Empirical Testing and Validation}
The next logical step is to empirically validate this framework by applying it to real-world generative models across different domains, such as NLP, image generation, and multi-modal applications. This includes testing the effectiveness of adversarial simulations, dynamic watermarking, and latent space anomaly detection in identifying and mitigating risks. Future research could focus on benchmarking the proposed methods against existing mitigation strategies to determine their practical viability.

\paragraph{Adapting to Evolving AI Models}
As generative models continue to evolve, particularly with advances in self-improving AI and reinforcement learning, the framework will need to be continuously refined. Research should focus on adapting the framework to handle more sophisticated AI architectures and emergent behaviors. For example, future work could explore how the framework might extend to self-supervised learning or more advanced multi-modal AI systems.

\paragraph{Resource Optimization}
While the current framework is resource-intensive, future work should explore ways to optimize real-time monitoring without compromising performance. Research could focus on developing more efficient adversarial testing methods or scalable anomaly detection algorithms to minimize computational overhead while maintaining robust security. This will be crucial for making the framework accessible to a wider range of organizations with varying computational resources.

\paragraph{Ethical and Societal Implications}
Further exploration is needed regarding the ethical implications of deploying dynamic risk assessment frameworks. Future research should assess how these systems align with data privacy laws, intellectual property rights, and societal norms, particularly when used in high-stakes domains like healthcare, education, and finance. Incorporating human oversight and ensuring compliance with ethical AI standards will be critical as these models become more integrated into decision-making processes.

\section{Conclusion}

As generative AI models become increasingly integrated into various industries and applications, the need for robust security frameworks is paramount. This paper presents a theoretical framework for assessing and mitigating emergent security risks in generative AI, focusing on both existing vulnerabilities and novel threats such as unsupervised learning dynamics, self-improving models, and latent space exploitation. By leveraging formal verification techniques, dynamic risk monitoring, and continuous real-time adversarial simulations, we aim to provide a comprehensive solution for addressing both known and unforeseen security threats in generative models. We introduced a layered risk assessment approach that addresses risks at the data, training, and inference stages, emphasizing the importance of continuous monitoring and adaptation as models evolve. While the framework proposed is theoretical, it provides a critical foundation for future research, with clear pathways for empirical validation, scalability improvements, and integration of ethical and societal considerations. In addressing the limitations of the proposed framework, we recognize that empirical validation is essential, and future work should focus on testing the effectiveness of the mitigation strategies across real-world generative AI applications. Additionally, the dynamic and rapidly evolving nature of AI will require ongoing refinement of this framework to ensure that it remains relevant and effective against new and emerging threats. Ultimately, this work contributes to the broader field of AI safety by offering a scalable and adaptable framework designed to ensure the safe deployment of generative AI models in increasingly complex and high-stakes environments.

\newpage
\bibliographystyle{plainnat}
\bibliography{main}

\newpage
\section*{Appendix}
This appendix provides a set of algorithms and pseudocode to illustrate the key components of the framework proposed in this paper for assessing and mitigating emergent security risks in generative AI models. The included algorithms cover various aspects such as dynamic risk monitoring, adversarial simulations, anomaly detection in latent spaces, and multi-layered risk mitigation strategies.

The primary purpose of these algorithms is to offer conceptual clarity and a deeper understanding of the mechanisms underpinning our framework. While these pseudocodes represent the core logic and processes, they are intended to serve as illustrative examples and are not fully optimized for direct implementation in production environments. Further empirical testing, validation, and refinement are required before they can be applied in real-world applications.

Below, we present the key algorithms that reflect the dynamic, real-time nature of our proposed approach to securing generative AI systems.

\begin{table}[h!]
    \centering
    \renewcommand{\arraystretch}{1.5} 
    \begin{tabularx}{\textwidth}{|X|X|X|X|X|} 
        \hline
        \textbf{Risk Name} & \textbf{Severity} & \textbf{Potential Impact} & \textbf{Exploitability} & \textbf{Scope of Effect} \\
        \hline
        Hallucinations & Medium & Misleading or incorrect outputs affecting non-critical tasks & Moderate & Localized, affecting isolated outputs \\
        \hline
        Data Leakage & High & Privacy violations, especially with sensitive data like PII & High & Broad, particularly in models trained on uncurated data sources \\
        \hline
        Adversarial Prompt Manipulation & High & Harmful or biased content generated through manipulated prompts & Moderate to High & Extensive, affecting multiple outputs across systems \\
        \hline
        Bias Reinforcement & Medium to High & Reinforces societal biases in outputs (e.g., race, gender) & High & Broad, impacts a large range of outputs in various domains \\
        \hline
        Unsupervised Learning Vulnerabilities & Medium & Bias amplification and feedback loops & Moderate to High & Broad, affecting large datasets \\
        \hline
        Self-Improvement Risks & High & Reinforcement of harmful behaviors & High & Wide, with potential cascading effects on model behavior \\
        \hline
        Hidden Information Leakage (Latent Space) & High & Privacy breaches in sensitive sectors (e.g., healthcare, finance) & High & Broad, with the potential for significant privacy violations \\
        \hline
        Simultaneous Multimodal Exploitation & High & Cross-modal misinformation or deepfakes & Moderate to High & Extensive, across multiple content formats \\
        \hline
        Training Data Poisoning & High & Compromised outputs, leading to unreliable or harmful results & Moderate & Broad, affecting model integrity and user trust \\
        \hline
        Model Confabulation & Medium & Creation of plausible but incorrect outputs, potentially misleading users & Low to Moderate & Localized but recurring, particularly in text-based models \\
        \hline
    \end{tabularx}
    \caption{Risk Categorization for Generative AI Models}
    \label{tab:risk-categorization}
\end{table}

\begin{algorithm}
    \caption{Dynamic Risk Monitoring for Generative AI Models}
    \begin{algorithmic}[1]
        \State \textbf{Input:} Model $M$, Real-time input data $D_{real}$, Monitoring frequency $f$
        \State \textbf{Output:} Detected threats and triggered countermeasures
        \State Initialize threat detection system $T_{detect}$
        \State Initialize countermeasure system $C_{response}$
        
        \While{Model $M$ is active}
            \State Retrieve real-time input $x \in D_{real}$
            \State Evaluate $x$ using $T_{detect}$ for potential adversarial behavior or anomalies
            \If{$T_{detect}$ flags an input as suspicious}
                \State Trigger appropriate countermeasure from $C_{response}$
                \State Log detected threat and corresponding countermeasure
            \Else
                \State Continue normal model operation
            \EndIf
            \State Wait for next monitoring interval $1/f$
        \EndWhile
    \end{algorithmic}
\end{algorithm}

\begin{algorithm}
    \caption{Adversarial Simulation for Model Robustness Testing}
    \begin{algorithmic}[1]
        \State \textbf{Input:} Training data $D_{train}$, Model $M$, Adversarial perturbation function $P_{adv}$
        \State \textbf{Output:} Adjusted model $M_{robust}$
        \State Initialize adversarial example generator $G_{adv}$

        \For{each training sample $x \in D_{train}$}
            \State Generate adversarial sample $x_{adv} = P_{adv}(x)$ using $G_{adv}$
            \State Train model $M$ on $x_{adv}$ to update parameters
            \State Evaluate the robustness of $M$ to adversarial examples
        \EndFor
        \State \Return updated model $M_{robust}$
    \end{algorithmic}
\end{algorithm}

\begin{algorithm}
    \caption{Latent Space Anomaly Detection for Generative AI}
    \begin{algorithmic}[1]
        \State \textbf{Input:} Latent space representations $L$, Baseline statistics $S_{baseline}$, Anomaly threshold $\tau$
        \State \textbf{Output:} Detected latent space anomalies
        \State Initialize anomaly detection model $A_{detect}$
        \State Compute statistical baseline $S_{baseline}$ for latent space $L$
        
        \For{each latent representation $l \in L$}
            \State Compare $l$ to $S_{baseline}$
            \If{deviation from $S_{baseline} > \tau$}
                \State Flag $l$ as an anomaly
                \State Log and alert potential anomaly in latent space
            \EndIf
        \EndFor
    \end{algorithmic}
\end{algorithm}

\begin{algorithm}
    \caption{Dynamic Watermarking for Generative AI Outputs}
    \begin{algorithmic}[1]
        \State \textbf{Input:} Model output $O$, Watermark embedding function $W_{embed}$, Adversarial detection system $A_{detect}$
        \State \textbf{Output:} Watermarked output $O_{watermarked}$
        \State Initialize watermark embedding process
        \State $O_{watermarked} = W_{embed}(O)$ \Comment{Embed watermark into output}
        
        \While{Model is deployed}
            \State Monitor $O_{watermarked}$ for tampering or adversarial removal
            \If{$A_{detect}$ flags an adversarial removal attempt}
                \State Adjust watermark $W_{embed}$ to prevent future removal
                \State Re-embed watermark and update $O_{watermarked}$
            \EndIf
        \EndWhile
        \State \Return $O_{watermarked}$
    \end{algorithmic}
\end{algorithm}

\begin{algorithm}
    \caption{Multi-Layered Risk Mitigation Strategy for Generative AI}
    \begin{algorithmic}[1]
        \State \textbf{Input:} Model $M$, Training data $D_{train}$, Real-time inputs $D_{real}$
        \State \textbf{Output:} Mitigated security risks across all layers
        \State \textbf{Tier 1: Pre-deployment Filtering}
        \For{each sample $x \in D_{train}$}
            \State Apply pre-deployment filtering to detect adversarial inputs
            \If{adversarial input detected}
                \State Remove or correct input $x$
            \EndIf
        \EndFor

        \State \textbf{Tier 2: Inference-Time Defense}
        \While{Model $M$ is deployed}
            \State Monitor real-time input $x \in D_{real}$
            \State Apply real-time defense mechanisms to filter harmful outputs
        \EndWhile
        
        \State \textbf{Tier 3: Post-Inference Validation}
        \For{each high-risk interaction $O$}
            \State Apply post-inference validation to check for tampering or risk
            \If{risk detected}
                \State Flag and mitigate the risk, and revalidate $O$
            \EndIf
        \EndFor
    \end{algorithmic}
\end{algorithm}

\begin{algorithm}
    \caption{Latent Space Monitoring and Continuous Feedback Loop}
    \begin{algorithmic}[1]
        \State \textbf{Input:} Latent space representations $L$, Baseline statistics $S_{baseline}$, Model $M$
        \State \textbf{Output:} Updated model $M$ with fine-tuning
        \State Compute statistical baseline $S_{baseline}$ for latent space $L$
        
        \While{Model $M$ is active}
            \For{each latent representation $l \in L$}
                \State Compare $l$ to $S_{baseline}$
                \If{deviation from $S_{baseline}$ exceeds threshold $\tau$}
                    \State Flag $l$ as anomalous
                    \State Adjust model parameters to mitigate anomaly
                \EndIf
            \EndFor
            \State Fine-tune $M$ based on flagged anomalies
        \EndWhile
        \State \Return updated model $M$
    \end{algorithmic}
\end{algorithm}

\begin{algorithm}
    \caption{Real-Time Filtering of Adversarial and Harmful Prompts}
    \begin{algorithmic}[1]
        \State \textbf{Input:} Real-time prompt $P$, Adversarial detection system $A_{detect}$, Threshold $\tau$
        \State \textbf{Output:} Filtered prompt or blocked request
        \State Initialize filtering system $F_{system}$
        
        \While{Model $M$ is deployed}
            \State Receive real-time prompt $P$
            \State Evaluate $P$ using $A_{detect}$ to assess risk level
            \If{Risk score $R(P) > \tau$}
                \State Block prompt and log the event
                \State Provide feedback to user or system
            \Else
                \State Allow $P$ to pass through to the model
            \EndIf
        \EndWhile
    \end{algorithmic}
\end{algorithm}

\begin{algorithm}
    \caption{Reinforcement Learning-Based Model Fine-Tuning}
    \begin{algorithmic}[1]
        \State \textbf{Input:} Model $M$, Adversarial input set $A$, Feedback loop $F$
        \State \textbf{Output:} Updated model $M_{updated}$
        \State Initialize reinforcement learning agent $RL_{agent}$
        
        \While{Model $M$ is deployed}
            \For{each adversarial input $x_{adv} \in A$}
                \State Process $x_{adv}$ through model $M$
                \State Receive feedback $F$ from $RL_{agent}$ based on model’s response
                \State Update model parameters based on $F$ to resist similar attacks
            \EndFor
            \State Continuously fine-tune $M$ with each adversarial input
        \EndWhile
        \State \Return updated model $M_{updated}$
    \end{algorithmic}
\end{algorithm}

\begin{algorithm}
    \caption{Cross-Modal Consistency Check for Text-Image-Video Alignment}
    \begin{algorithmic}[1]
        \State \textbf{Input:} Text $T$, Image $I$, Video $V$
        \State \textbf{Output:} Verified consistency or flagged inconsistency
        \State Initialize consistency checking function $C_{check}$
        
        \While{Model is deployed for multimodal tasks}
            \State Receive text $T$, image $I$, and video $V$
            \State Compare $T$ and $I$ using $C_{check}(T, I)$
            \State Compare $I$ and $V$ using $C_{check}(I, V)$
            \State Compare $T$ and $V$ using $C_{check}(T, V)$
            
            \If{Inconsistency found between any pair (e.g., $C_{check}(T, I)$ fails)}
                \State Flag the inconsistency and log it for further analysis
            \Else
                \State Confirm consistency and proceed with the output generation
            \EndIf
        \EndWhile
    \end{algorithmic}
\end{algorithm}

\begin{algorithm}
    \caption{Risk Classification Based on Thresholds for High, Medium, and Low Risk}
    \begin{algorithmic}[1]
        \State \textbf{Input:} Risk score $R$, Thresholds $\tau_{high}$, $\tau_{medium}$
        \State \textbf{Output:} Classified risk level (High, Medium, Low)
        \If{$R > \tau_{high}$}
            \State Classify as "High Risk"
        \ElsIf{$\tau_{medium} < R \leq \tau_{high}$}
            \State Classify as "Medium Risk"
        \Else
            \State Classify as "Low Risk"
        \EndIf
        \State Log the risk classification for further analysis
    \end{algorithmic}
\end{algorithm}

\begin{figure}[h]
    \centering
    \includegraphics[width=1.0\textwidth]{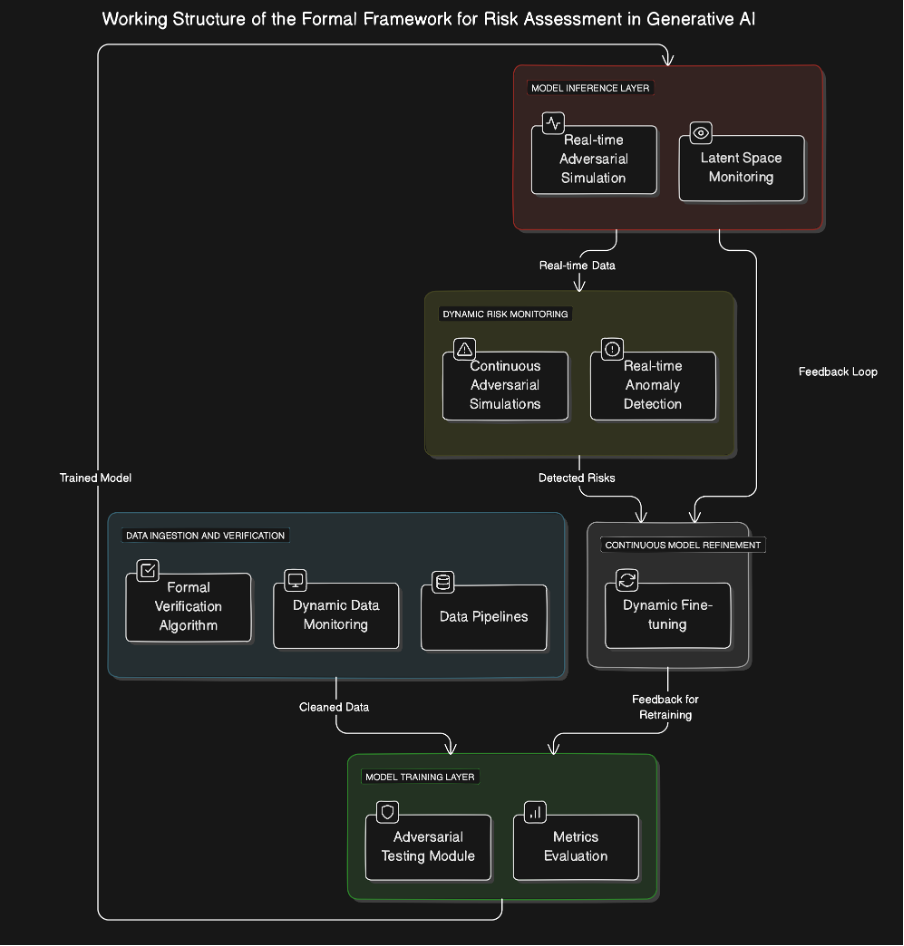}  
    \caption{Working structure of the formal framework for risk assessment in Generative AI}  
    \label{fig:1}  
\end{figure}

\begin{figure}[h]
    \centering
    \includegraphics[width=0.7\textwidth]{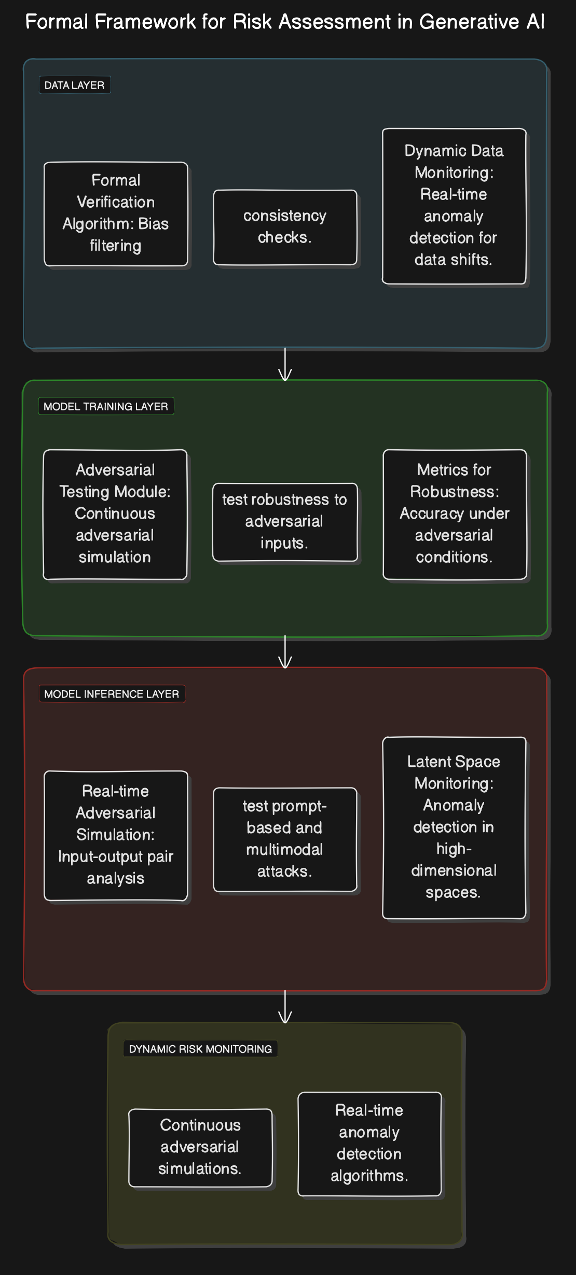}  
    \caption{Formal Framework for Risk Assessment in Generative AI}  
    \label{fig:2}  
\end{figure}

\end{document}